# Antimatter in the direct-action theory of fields

Ruth E. Kastner

13 September 2015

ABSTRACT. One of Feynman's greatest contributions to physics was the interpretation of negative energies as antimatter in quantum field theory. A key component of this interpretation is the Feynman propagator, which seeks to describe the behavior of antimatter at the virtual particle level. Ironically, it turns out that one can dispense with the Feynman propagator in a direct-action theory of fields, while still retaining the interpretation of negative energy solutions as antiparticles.

1. Introduction

This issue salutes the profound contributions by Richard P. Feynman. Feynman is known for the Wheeler-Feynman (W-F) 'direct-action' theory of classical electromagnetism [1]. Another of his contributions is the interpretation of the negative energy field equation solutions as antiparticles, and the invention of the 'Feynman propagator' which incorporates the antiparticle concept into virtual propagation. In this paper, I examine these key features of Feynman's work and attempt to elucidate their relationship to my recent development of the Transactional Interpretation (TI) of quantum theory, which is based on the W-F direct-action theory. The first three sections are review, while sections 4 and 5 are original research.

John G. Cramer used the Wheeler-Feynman theory as the basis of TI [2]. I have proposed a relativistic extension of TI [3] based on the direct-action theory of QED elaborated by Davies [4]. I call this relativistic version of TI the 'Possibilist Transactional Interpretation' (PTI), because the quantum states are interpreted as extra-spatiotemporal possibilities. (For the specific details of this suggested ontology, the reader is invited to consult [3], Chapter 7.) In the Davies theory, and accordingly in PTI, virtual particle processes are described not by the Feynman propagator, but by the time-symmetric propagator. The question then naturally arises: what exactly is an antiparticle in PTI? This paper will addressed that question, as well as the historically curious fact that Feynman abandoned his own direct-action theory. In what follows, I will first briefly consider the latter historical issue and then, in some depth, the former theoretical one.

2. Feynman's unnecessary abandonment of the Wheeler-Feynman direct-action theory

Feynman's primary concern was the infinite energy of self-action plaguing classical electromagnetism. In classical electromagnetism, energy is straightforwardly carried by the field, and thus for Feynman the key aspect of the direct-action theory was the restriction that a charge could not interact with its own field. Instead, the field acting on a given particle was due to the advanced responses of absorbers to the emitted field of the particle. This provided an elegant account of radiative damping (i.e. the loss of energy by a radiating charge).

But when Feynman turned his attention to the quantum level, he noted that certain relativistic effects, such as the Lamb shift, required some form of self-action of a charge. Unable to reconcile this with his assumption that self-action cannot be allowed in a direct-action theory, he abandoned it. However, that assumption was apparently not correct. It was based on the idea that all interactions involve energy transfer, which is not the case at the quantum level.

In the early 1970's, Davies plunged onward, and provided a fully developed quantum version of the direct-action theory [4]. This theory incorporates the fact that, at the quantum level, currents are indistinguishable, so there is no way to say whether or not a given current is undergoing self-action. However, this does not result in infinities, since the self-action does not lead to energy transfer from a current to itself. This is because, as noted in Section 1 above, those self-action processes are at the virtual level only. The virtual interaction does not prompt an advanced response that provides for the radiation of energy from an emitting current. (See also [5] for a discussion of this point.)

In mathematical terms, the virtual processes are described by the time-symmetric propagator, which cannot transfer energy. Energy is only transferred in the presence of advanced responses from absorbing currents. Specifically, Davies shows that if all emitted fields are absorbed, the responses of absorbers together with the basic time-symmetric propagation yields precisely the Feynman propagator. Davies notes that the Feynman propagator $D_F$ can be decomposed as follows:

$$D_F = \overline{D} + \frac{1}{2}(D^+ - D^-) \tag{1}$$

where $\overline{D}$ is the time-symmetric propagator, and the quantity in parentheses is the difference of the positive and negative frequency components of the solution to the homogeneous field equation. Davies notes that the first term $\overline{D}$ is just the basic time-symmetric direct interaction between currents. Taking into account the responses of all the currents and the relationship

$$D^+(x-y) = -D^-(y-x) \tag{2}$$

and integrating over all spacetime coordinates transforms the second term of (1) into only positive frequency free field components, i.e. $D^+(x-y)$. For the universe as a whole, this contribution must vanish since the 'light tight box' condition precludes the existence of any nonzero free field. But for a subset of currents constituting some system of interest, this term represents the emission of a real photon by the system, and this is how energy is transferred from one system to another. The interesting point is that the energy transferred through the direct-action theory based on the responses of absorbers is always positive; this is because of the relation (2) and the fact that all currents are symmetrically summed over (see [3], Davies [1971], pp. 840-843 for details).[1] Thus the negative frequency field solutions are duly taken into account, but always result in the transfer of positive energy in any empirical process.

Energy conservation precludes a current from responding to its own emitted field, and thus at the quantum level one can have self-action at the field level through the time-symmetric component $\overline{D}$ without any corresponding energy infinities. (I have noted elsewhere [6] that this ability of the direct-action theory to avoid self-action infinities makes it an excellent candidate for resolving the notorious consistency problems of standard quantum field theories.)

Though Feynman abandoned his direct-action theory, it was recently enthusiastically resurrected by his co-originator John Wheeler [7]. In a paper with D. Wesley, Wheeler commented that

> [WF] swept the electromagnetic field from between the charged particles and replaced it with "half-retarded, half advanced direct interaction" between particle and particle. It was the high point of this work to show that the standard and well-tested force of reaction of radiation on an accelerated charge is accounted for as the sum of the direct actions on that charge by all the charges of any distant complete absorber. Such a

---

[1] Summation over all spacetime indices is also an indication that this is not a spacetime process—i.e. that no particular region(s) of spacetime is/are involved.

formulation enforces global physical laws, and results in a quantitatively correct description of radiative phenomena, without assigning stress-energy to the electromagnetic field. ([6], p. 427)

Thus it is clear that the direct-action theory of fields is perfectly viable, despite Feynman's abandonment of it. In what follows, I will discuss how antimatter is treated in the direct action theory, and why Feynman's treatment of antiparticles via the Feynman propagator can be seen as 'overkill' in this context.

3. Antiparticles in relativistic quantum mechanics: a brief review

Relativistic wave equations have solutions characterized by both positive and negative energies. Consider, for example, the Dirac equation (in covariant notation):

$$(i\gamma^\mu \partial_\mu - m)\psi = 0 \qquad (3)$$

This describes fermions (such as the electron) and has four spinor solutions:

$$\psi_i = u_i(p_\mu)e^{-ip_\mu x^\mu} \qquad (4)$$

The spinors $u_i$ are 4-component vector states. For simplicity, consider the solutions for the frame in which the fermion is at rest ($\mathbf{p}=0$). Then the solutions $\psi_i$ are given by:

$$\psi_1 = \begin{pmatrix} 1 \\ 0 \\ 0 \\ 0 \end{pmatrix} e^{-imt} ; \quad \psi_2 = \begin{pmatrix} 0 \\ 1 \\ 0 \\ 0 \end{pmatrix} e^{-imt} ; \quad \psi_3 = \begin{pmatrix} 0 \\ 0 \\ 1 \\ 0 \end{pmatrix} e^{+imt} ; \quad \psi_4 = \begin{pmatrix} 0 \\ 0 \\ 0 \\ 1 \end{pmatrix} e^{+imt} \qquad (5)$$

Solutions $\psi_1$ and $\psi_2$ describe positive energies, E= m, while solutions $\psi_3$ and $\psi_4$ yield negative energies, E = −m. We need all four solutions for a complete set of solutions to the equation. Thus, even a first-order equation such as the Dirac equation forces negative energies that cannot be simply thrown out as 'unphysical.' In what follows, we'll work with the (simpler) quantized solutions to the Klein-Gordon equation to further illustrate how these negative energies are encountered and interpreted as antiparticles. This standard interpretation goes over into PTI, since the radiation field is effectively quantized in that theory (the Coulomb field is not). In PTI, the quantization enters in through the transactional process, in which an offer wave (excited state of the field) is created and responded to by its adjoint (the confirmation from an absorber).[2] We will see that the only difference between the usual treatment of antiparticles is at the virtual level, involving propagators.

Here we wlll roughly follow the pedagogically clear account in Teller [8]. In what is often termed 'second quantization', the solution $\Psi(\vec{x},t)$ of a wave equation is promoted to an operator $\hat{\Psi}(\vec{x},t)$ and Fourier analyzed in terms of frequencies of component quantized oscillators, i.e.[3]

$$\hat{\Psi}(\vec{x},t) = \int d^3\widetilde{k}\, \hat{a}(\vec{k},t) e^{i\vec{k}\cdot\vec{x}} \qquad (6)$$

where the $\hat{a}(\vec{k},t)$ are annihilation operators and their adjoints $\hat{a}^\dagger(\vec{k},t)$ are creation operators. For the Klein-Gordon equation (and indeed any relativistic wave equation), the operator coefficients $\hat{a}(\vec{k},t)$ can have not only the usual retarded time dependence,

$$\hat{a}(\vec{k},t)_{ret} = \hat{a}(\vec{k}) e^{-i\omega t} \qquad (7)$$

but also the advanced time dependence

$$\hat{a}(\vec{k},t)_{adv} = \hat{a}(\vec{k}) e^{+i\omega t}. \qquad (8)$$

Now, each oscillator can be described by a generalized real-valued coordinate $\hat{q}(k,t)$:

$$\hat{q}(\vec{k},t) \propto \hat{a}(\vec{k},t) + \hat{a}^\dagger(\vec{k},t) \qquad (9)$$

whose momentum is $\hat{p} = \partial_t \hat{q}(\vec{k},t)$.

For the usual retarded solutions (5) we get the following expression for momentum:

$$\hat{p}_{ret}(\vec{k},t) \propto -i[\hat{a}(\vec{k},t) - \hat{a}^\dagger(\vec{k},t)] \qquad (10)$$

---

[2] This is the case despite the fact that PTI is based on a direct-action theory of fields (i.e., Davies [4]). The basic field propagation is via a nonlocal connection between currents; this is virtual propagation. However, when there is a well-defined emission and absorption (see [5]), the result is the transfer of a photon that formally corresponds to a Fock state. That state can be represented by the action of a creation operator on the vacuum.

[3] Following Teller [8], normalization is included in the integration variable $\vec{k}$, indicated by the tilde.

However, for the advanced ('negative energy') solutions (6) we get a momentum with the opposite sign, i.e.:

$$\hat{p}_{adv}(\vec{k},t) \propto i[\hat{a}(\vec{k},t) - \hat{a}^\dagger(\vec{k},t)] = -\hat{p}_{ret}(\vec{k},t) \qquad (11)$$

Now, recalling that momentum is the generator of spatial translations, (11) has the inconsistent property that it associates a *negative* spatial translation with a *positive* wave vector $\vec{k}$. This is rectified by interchanging the roles of the creation and annihilation operators, thus defining a new set whose arguments have the opposite sign for $\vec{k}$, i.e.:

$$\hat{a}(\vec{k},t)_{adv} = \hat{b}^\dagger(-\vec{k})e^{+i\omega t} \qquad (12a)$$

$$\hat{a}^\dagger(\vec{k},t)_{adv} = \hat{b}(-\vec{k})e^{-i\omega t} \qquad (12b)$$

We now generalize the basic field expression (6) to explicitly include both the retarded and advanced solutions, the latter being re-expressed with the sign of $\vec{k}$ flipped. Thus we have:

$$\hat{\Psi}(\vec{x},t) = \int d^3\widetilde{k}[\hat{a}(\vec{k})e^{i(\vec{k}\cdot\vec{x}-\omega t)} + \hat{b}^\dagger(-\vec{k})e^{i(\vec{k}\cdot\vec{x}+\omega t)}] = \int d^3\widetilde{k}[\hat{a}(\vec{k})e^{i(\vec{k}\cdot\vec{x}-\omega t)} + \hat{b}^\dagger(\vec{k})e^{-i(\vec{k}\cdot\vec{x}-\omega t)}]$$
$$(13)$$

We can see that having restored a consistent account of the oscillator momentum $p$, the 'advanced' solutions no longer appear advanced. However, technically they still carry negative energy eigenvalues, since

$$i\frac{d}{dt}\hat{b}^\dagger(\vec{k},t) = -\omega\hat{b}^\dagger(\vec{k},t) \qquad (14)$$

One now interprets the $b$ operators as creation and annihilation operators of *antiparticles* with positive energy, yet with features opposite to their respective particles. Heuristically, we can think of the antiparticles as 'holes' that behave as inverse counterparts of their corresponding quanta: Equation (12(a)) describes the removal of a quantum of momentum $\vec{k}$ (i.e., resulting in a 'hole') as equivalent to *adding* an antiquantum with opposite momentum $-\vec{k}$ (creating an antiquantum). Along with the above reinterpretation defining antiparticles, one redefines the energy operator with a flipped sign to reflect the fact that the physically observed energy of the antiparticles is positive.

To gain some physical insight into the idea that antiparticles carry positive energy, consider the analogy with monetary gain or loss. We can gain money in two ways: (1) receiving a payment, or (2) having a debt forgiven. (1) corresponds to adding a quantum of positive energy and (2) corresponds to removing a quantum of negative energy. But in order to effect (2), we still need to *receive* something—what we receive is a (2') notice of cancellation of the debt. The object (2') corresponds to the antiquantum with positive energy.

4. Antiparticles in PTI

So how does all this relate to PTI? In PTI, an offer wave can be represented by a Fock state, which can be obtained as the action of a creation operator on the vacuum state,[4] i.e.

$$\hat{b}^\dagger(\vec{k})|0\rangle = |\overline{\vec{k}}\rangle \qquad (15)$$

where the bar indicates an antiparticle state in this case.
The generation of a confirmation wave can be represented by the annihilation operator acting on the adjoint vacuum, i.e.

$$\langle 0|\hat{b}(\vec{k}) = \langle\overline{\vec{k}}| \qquad (16)$$

The only difference between the standard account of the fields and that of PTI is that in the latter, the integral over spatial momentum $\vec{k}$ in (13) becomes a discrete sum. This is because, at the relativistic level, emission of an offer wave $|\vec{k}\rangle$ is contingent on there being an absorber eligible to respond to that particular momentum value. Discreteness in absorber availability (i.e. the fact that excitable quantum systems such as atoms or molecules do not form a continuum) dictates that directional momentum components do not obtain as a continuum in the emitted offer wave.

Thus, we see that antiparticles, considered as "real quanta" (offers and confirmations leading to transactions), go over in essentially the standard way to PTI.[5] The main difference between PTI and the usual account of antiparticles enters at the virtual particle level. We turn to this issue in the next section.

---

[4] In a standard direct-action theory, creation and destruction operators do not appear because the field is nonquantized. However, in PTI the radiation component of the field (but not the Coulomb component) is quantized, since actualized transactions are identified with radiated, 'real' photons. (cf. Kastner [5]) This interpretation differs from that of Davies [4] who makes no fundamental distinction between Coulomb field and the radiation field, and does not apply a transactional interpretation to his theory. But it also aligns nicely with that of Rohrlich [9] who proposed that the radiation field be quantized and not the Coulomb field, to resolve the problem of self-action.

5. Virtual particles: the standard account vs. the PTI account.

First, for convenience we write the full expression for the field along with its adjoint, since both are needed in the general propagator (virtual particle) expressions involving charged fields:

$$\hat{\Psi}(\vec{x},t) = \int d^3\tilde{k}[\hat{a}(\vec{k})e^{i(\vec{k}\cdot\vec{x}-\omega t)} + \hat{b}^\dagger(\vec{k})e^{-i(\vec{k}\cdot\vec{x}-\omega t)}]$$

$$\hat{\Psi}^\dagger(\vec{x},t) = \int d^3\tilde{k}[\hat{a}^\dagger(\vec{k})e^{-i(\vec{k}\cdot\vec{x}-\omega t)} + \hat{b}(\vec{k})e^{i(\vec{k}\cdot\vec{x}-\omega t)}]$$

(17)

The Feynman propagator $D_F$ can be written as a time-ordered 'vacuum expectation value' of fields, i.e:

$$iD_F(x-y) = T(\langle 0|\Psi(x)\Psi^\dagger(y)|0\rangle)$$

(18)

The notation on the right-hand side specifies that for $x^0 < y^0$ the product is carried out as written; but for $y^0 < x^0$, the order of the field operators is reversed, i.e. : $\langle 0|\Psi^\dagger(x)\Psi(y)|0\rangle$
This is done in order to attempt to 'enforce causality', i.e., make sure that no quantum is being destroyed before it is created. This is viewed as necessary because at the level of virtual propagation, there is no restriction on the locations in spacetime of the site of emission $x$ and the site of absorption $y$ (where these are coordinate 4-vectors); so, in principle, these could be spacelike separated. For such cases, there is no fact of the matter about whether $x^0 < y^0$ or vice versa; the order is frame-dependent. So the intent of the Feynman propagator is to ensure that in a frame in which $x^0 < y^0$, a particle is emitted at x and absorbed at y; but in a frame for which $y^0 < x^0$, an antiparticle is emitted at y and absorbed at x. A clear account of the requirement for the Feynman propagator under the assumption that all propagation is contained in spacetime, along with a critique of how the issue of causality is handled in QFT, is presented in [12].

---

[5] QFT does face an ambiguity in curved spacetime with respect to the division into particle and antiparticle field operators, as noted by Hawking [11]. This may present a consistency issue for the Feynman propagator, since the latter implicitly invokes a Fock space state that may not be well-defined under these conditions (see §5, (19)). In contrast, PTI does not need a globally well-defined Fock space, since the quantized component is contingent on specific absorber responses, and therefore does not need to be well defined independently of such responses.

Now, this is a fine way to do careful bookkeeping if one assumes that all these processes are occurring in spacetime.[6] However, in PTI the assumption that quantum processes take place in spacetime is relinquished. In addition, in PTI the basic field propagation is time-symmetric. This is reflected in the description of virtual processes by the time-symmetric propagator (rather than the Feynman propagator). In more fundamental ontological terms, it is *energy-symmetric*: there is no restriction placed on the propagation of either positive or negative energies. The interaction represented by the energy-symmetric propagator involves neither a real emission of an offer wave nor any absorber response (in contrast to the cases of real quanta discussed in the previous section). Thus, the energy-symmetric propagator properly takes into account the natural ambiguity of causal direction in a process which, by its very nature, is pre-causal. The propagator in PTI is a direct connection between currents; it is a pre-spacetime process and as such can have no temporal direction.

Recall that virtual particles (whose behavior is represented by the propagator) are off the mass shell, i.e., they do not satisfy the relativistic constraint $\omega^2 - \vec{k}^2 = m^2$ and thus are not subject to the limitation of light speed. Given the strange properties of virtual particles, PTI embraces the idea that their behaviors are not spacetime phenomena, and this allows us to dispense with the bookkeeping involved in making sure that 'positive energy goes forward in time' and 'negative energy goes backward in time' that is the essence of the Feynman propagator. In fact, virtual particles are not going *anywhere/when* in space or time. They are neither real particles nor real antiparticles; they are never detected or detectable, and are thus purely sub-empirical. In this picture, the attempt to 'enforce causality' is inappropriate at the level of virtual particle propagation; it is a holdover from the unnecessary assumption that all processes must be spacetime processes.

Another way to see why the direct-action theory does not need the bookkeeping involved in the Feynman propagator is to consider again the interesting fact that due to relation (2), the negative frequencies $D^-(x-y)$ rear their 'ugly heads' but then vanish without making any empirical contribution. Moreover there is no need to distinguish between the case $D^+(x-y)$ and $D^+(y-x)$ because the double summation over all currents and all spacetime coordinates (which in PTI are just *possible* spacetime locations) renders them equivalent. Thus the Feynman propagator, which makes a distinction between the two cases, is not needed in the direct-action picture.[7]

---

[6] This author is aware of the highly nontrivial ontological considerations surrounding the nature of a 'process' that is not a spacetime process. Those issues are beyond the scope of this paper, but it should be noted here that they may be understood in a Whiteheadian light, e.g. wherein the "actual [spacetime] entity is the real concrescence of many [pre-spacetime] potentials." [14], "Categories of Explanation."

Davies [4] further reminds us that the Feynman propagator can be decomposed into a real time/energy-symmetric part and an imaginary singular part:

$$D_F(x) = \frac{1}{(2\pi)^4} \int \left( \frac{PP}{k^2} - i\pi\delta(k^2) \right) e^{ikx} dk = \overline{D} + D_1 \qquad (19)$$

where 'PP' signifies the principal part and $D_1$ is the vacuum expectation value of the anticommutator of two field operators. This latter term becomes equivalent to $D^+$ in the direct-action picture when integrated over all spacetime coordinates and all currents are summed over, as discussed above in §2. In this form, it can be seen that the Feynman propagator ambiguously covers two different physical situations: off-shell virtual quanta (described by the real, time-symmetric principal part) and real quanta (on-shell, described by the delta function term). In contrast, PTI makes a clean distinction between these two situations: virtual quanta are described only by $\overline{D}$ and real quanta are described only by $D_1$.

One might still ask: what about the nice picture of antiquanta as time-reversed counterparts of their respective quanta, which appears so clearly in Feynman diagrams? This intuitive understanding can be retained for the real quanta, which in TI correspond to the probability current.[8] Considering the probability 4-current for the Dirac field, the negative energy solutions have a positive-definite probability density (zeroth component of the 4-current). However, the probability flux (spatial components 1-3) proceeds in the opposite direction from the negative energy state's momentum. This fits very nicely the intuitive picture we see in Feynman diagrams, in which the antiquanta behave oppositely from their respective quanta.

6. Conclusion

---

[7] We are working with several different fields in this paper, but the basic principles presented in terms of one type of field hold for the analogous quantities (i.e., the propagators and field operators) in the other fields.
[8] There is an ambiguity surrounding the term 'real quantum' due to the inherent ambiguity of standard QFT, which does not distinguish between (i) a quantum state and (ii) an empirically detected quantum. In PTI, (i) is the offer wave and (ii) is the weighted projection operator that includes the confirming response to the offer (cf. [3], p. 54 and p.132). In this context we are referring to (ii).

It has been noted that relativistic wave equations mandate that negative energy solutions be included, and that their natural interpretation is as antiparticles. The standard quantum field theoretic account has been reviewed, with the observation that the results transfer immediately to the possibilist transactional interpretation (PTI) for transacted real quanta, which can be viewed as field excitations. In contrast however, there is no need for the time-ordered Feynman propagator to describe virtual quanta in PTI, since virtual quanta are unambiguously described by the time-symmetric propagator.

The Feynman propagator is recovered in the direct-action theory upon which PTI is based when absorber response is added to the time-symmetric propagator. It then functions as an ambiguous entity that describes both virtual particle propagation and real particle propagation. Since (in contrast to standard quantum field theory) PTI clearly distinguishes between these two, the Feynman propagator is not needed in this picture. One has *either* a time-symmetric virtual quantum (described by the time-symmetric propagator), *or* a Fock space state (a pole in the complex frequency plane).

Finally, it is interesting to note that Cohen and Elitzur [13] have observed in the context of 'interaction-free measurements' that "in any interaction of a particle with more than one possible absorber, each absorber's capability of absorbing the particle takes part in determining its final position ('collapse')". This observation is harmonious with the interpretation discussed here, although it should be emphasized that in PTI the actualized outcome is genuinely indeterministic. The 'paths not taken' (corresponding to the absorbers that do not receive the particle) are incipient transactions that were not actualized.

Acknowledgments.
I would like to thank Fred Alan Wolf for valuable correspondence and two anonymous referees for helpful suggestions for improvement of the presentation.